\documentclass[sort&compress
,final            
]{aipproc}

\layoutstyle{8x11double}

\usepackage{color}
\usepackage{nicefrac}

\begin{document}

\title{Re--thinking the Rubric for Grading the CUE:\\ The Superposition Principle}



\author{Justyna P. Zwolak}{address={Department of Physics, Oregon State University, Corvallis, OR 97331}}

\author{Mary Bridget Kustusch}{address={Department of Physics, Oregon State University, Corvallis, OR 97331}}

\author{Corinne A. Manogue}{address={Department of Physics, Oregon State University, Corvallis, OR 97331}}

\begin{abstract}
While introductory electricity and magnetism (E\&M) has been investigated for decades, research at the upper-division is relatively new. The University of Colorado has developed the Colorado Upper-Division Electrostatics (CUE) Diagnostic to test students' understanding of the content of the first semester of an upper-division E\&M course. While the questions on the CUE cover many learning goals in an appropriate manner, we believe the rubric for the CUE is particularly aligned to the topics and methods of teaching at the University of Colorado. We suggest that changes to the rubric would allow for better assessment of a wider range of teaching schemes. As an example, we highlight one problem from the CUE involving the superposition principle. Using data from both Oregon State University and the University of Colorado, we discuss the limitations of the current rubric, compare results using a different analysis scheme, and discuss the implications for assessing students' understanding.
\end{abstract}

\maketitle


\section{Introduction}

Introductory-level electricity and magnetism (E\&M) has been investigated since the early 80's (see references in \citet{McDermott99-RL}). Yet, research at the upper-division level is relatively new. Recently, the University of Colorado developed the Colorado Upper-Division Electrostatics (CUE) Diagnostic.\cite{Chasteen08-CUE,Chasteen09-CUE1,Pepper12-SDM,Chasteen12} One of the primary purposes of the CUE is to ``serve as a comparative instrument to assess upper-division E\&M courses.'' \cite{Chasteen09-CUE2}

The CUE\cite{CUEMI} is designed in a pre/post format. The optional 20-minute pre-test contains 7 out of the 17 post CUE questions which junior-level students might reasonably be expected to be able to solve based on their introductory course experience. The post-test is designed to be given at the end of the first upper-division semester in a single 50-minute lecture. The CUE contains open-ended questions where students do not actually solve the given problems. Instead, the instructions for the first half of the post-test are as follows:
\begin{quote}
For each of the following, give a brief outline of the EASIEST method that you would use to solve the problem. Methods used in this class include but are not limited to: Direct Integration, Ampere's Law, Superposition, Gauss' Law, Method of Images, Separation of Variables, and Multipole Expansion.

\end{quote}

\begin{figure}
\includegraphics[scale=0.75] {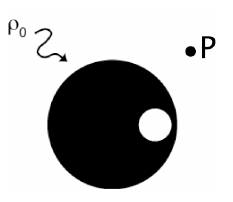}
 \caption{Question 5 reproduced from the CUE: ``A charged insulating solid sphere with a uniform volume charge density $\rho_0$, with an off-center spherical cavity carved out of it. Find E (or V) at point P, a distance 4R from the sphere.''\cite{CUEMI}}
\label{fig:q5}
\end{figure}

\begin{table}
\begin{tabular}{p{2.1cm}p{1cm}p{11.5cm}}
\hline
{\bf Correct} Answer &	3 points	&Correct answer is superposition\\
&&{\bf 0} points for only saying Gauss' Law.  \\
&&{\bf+1} point for saying integration or dipole\\
&&{\bf+1} point for superposition of charges but not fields (\emph{e.g.,} for $4/3\pi(R^3-r^3)\rho_0$)\\
&&0 for ``total charge of sphere with cavity''\\
\hline
{\bf Explanation}&	2 points	&Full answer is superposition of two oppositely charged spheres and then Gauss' Law to solve for E of each sphere.  Need to indicate what is being superposed for full credit \qquad (\emph{e.g.,} an antisphere of negative charge density).   \\
&&{\bf +1} point for stating what is superposed --- two oppositely charged sphere \\
&&({\bf +0.5} point if they don't state the spheres are oppositely charged)\\
&&{\bf +1} point for explaining how to solve using the two charged spheres\\
\hline
\end{tabular}
\caption{Grading rubric for CUE Question 5. Reproduced from the CUE Assessment Grading Rubric.\cite{CUEMI}}
\label{tab:q2/q5}
\end{table}

At Oregon State University, we have been collecting CUE data since 2009 and are now beginning to analyze it.  
As part of the Paradigms in Physics project, our curriculum was significantly restructured \cite{Manogue01-TUC,Manogue03-RU} and we introduced many novel active engagement strategies.\cite{wiki} As a result, we believe  a comparison between our students and those at Colorado will provide interesting information about how the CUE will generalize beyond the context at Colorado.

We have found that although the questions on the CUE reflect many of our learning goals in an appropriate manner, the current rubric for the CUE is more particularly aligned to the topics and methods of teaching at the University of Colorado. We suggest that changes to the rubric would allow for better assessment of a wider range of teaching schemes. As an example, this paper focuses on one problem from the CUE involving the superposition principle (shown in Fig.~\ref{fig:q5}). We discuss some limitations of the current rubric (shown in Table~\ref{tab:q2/q5}) and propose an alternative scheme for analyzing student responses to the problem. Using data from both Oregon State University (OSU) and the University of Colorado (CU), we show results with the new scheme and discuss implications for assessing students' understanding.

\section{Upper-division E\&M: traditional courses vs. Paradigms in physics}

\begin{figure}[b]
 \includegraphics[width=0.45\textwidth]{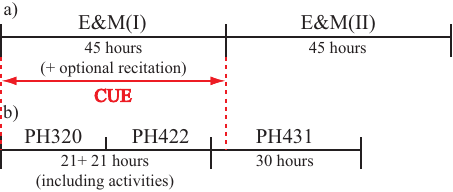}
 \caption{a) Standard semester schedule for E\&M courses. b) OSU quarter schedule for E\&M courses. }
\label{fig:emscheme}
\end{figure}

Traditionally, a first one-semester (15-16 weeks) E\&M~I course at a research university covers, for example, approximately the first six chapters of the text ``Introduction to Electrodynamics" by David J. Griffiths,\cite{GEM} \textit{i.e.,}  a review of the vector calculus necessary for a mathematical approach to the electricity and magnetism, electrostatics and magnetostatics in both vacuum and in matter.\cite{Chasteen12}  A second  semester course on Electrodynamics would typically cover most of the remaining chapters of Griffiths.

At Oregon State University, the middle-- and upper--division curriculum was extensively reordered compared to traditionally taught courses.\cite{Manogue01-TUC,Manogue03-RU} The first two Paradigms in Physics courses (PH 320: Symmetries and Idealizations and  PH 422: Static Vector Fields) cover electro- and magnetostatics in vacuum, approximately the material covered in Griffiths Chapters 1, 2 and 5. However,  we start with electrostatic potential (before electric fields) and integrate the mathematical methods with the physics content, including a strong emphasis on off-axis problems and power series approximations.\cite{wiki,LenThesis,GoldtrapThesis}  The remaining content of the standard E\&M~I curriculum is covered at the beginning of the senior year, as a part of PH 431 Capstones in Physics: Electromagnetism, which also covers some of the content of a more traditional E\&M~II course. 

Given the different course structure at OSU, not everything that the CUE tests is covered by the end of the fall quarter (Fig.~\ref{fig:emscheme}). In order to examine students' understanding of the content of PH 320 and PH 422, we selected a subset of 12 out of 17 CUE questions, which we give as a mid-test to OSU students at the end of the fall quarter of the junior year. The same mid-test is given again at the beginning of the senior year, as a part of PH 431, and the full CUE post-test is given at the end of the fall quarter of the senior year. 
In order to provide a reasonable comparison between CU and OSU, we compare CU post-test results on the question in Fig.~\ref{fig:q5} to results from OSU on the same question, given as part of the first mid-test at the end of the fall term in the junior year.

\section{Difficulties with the Rubric}

During an initial grading of OSU students using the rubric provided for this question (see Table~\ref{tab:q2/q5}), we noticed many solutions, including ones we would view as correct, did not seem to fit the rubric. In particular, OSU students often did not use the word ``superposition,'' instead trying to explain what they would do. Also, despite the problem statement explicitly allowing for a potential approach, this approach was absent in the rubric. 

More importantly, it was often not clear from students' answers what they wanted to add/superimpose --- fields, charges or something else --- even when they used the word ``superposition.'' Although the rubric accounts for situations where a student explicitly tries to superpose charges instead of fields, the ambiguous response is not accounted for in the rubric. Below are three examples, from the exams used for calibration, where the response is ambiguous about what is being superposed and yet each received almost full credit:
\begin{quote}
{\em ``Use superposition and have a normal sphere w/ charge density $\rho_0$ and a small sphere placed at $+x$ {\it (student marked $+x$ at the center of the cavity)} w/ charge density $-\rho_0$, superimpose one over the other.''} (Test 10: 5 points)
\end{quote}
\begin{quote}
{\em ``I might try some type of superposition here. It could be easy to subtract the off-centered spherical cavity.''} (Test 11: 4 points)
\end{quote}
\begin{quote}
{\em ``This would be solved using Gauss' law with the law of superposition to subtract the one sphere from the other.''} (Test 12: 4 points)
\end{quote}

To address these concerns, we developed a new categorization of responses for this question, shown in Table~\ref{tab:newq2/q5}, which focuses primarily on \textit{what} is being superposed and secondarily on whether the word ``superposition'' is used. In the following section, we present a comparison of OSU and CU using these new categories.

\begin{figure}[b]
 \includegraphics[height=0.21\textheight]{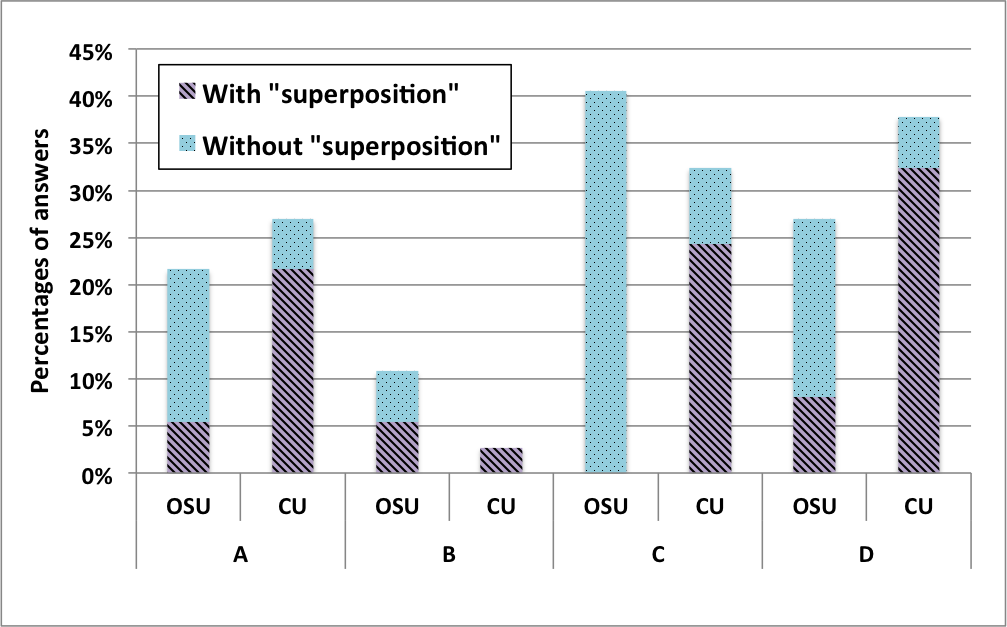}
 \caption{Frequency of use of the term ``superposition" in students' answers at OSU vs. CU (purple hatched pattern)}
\label{fig:q5data-rel}
\end{figure}

\begin{figure}[b]
 \includegraphics[height=0.21\textheight]{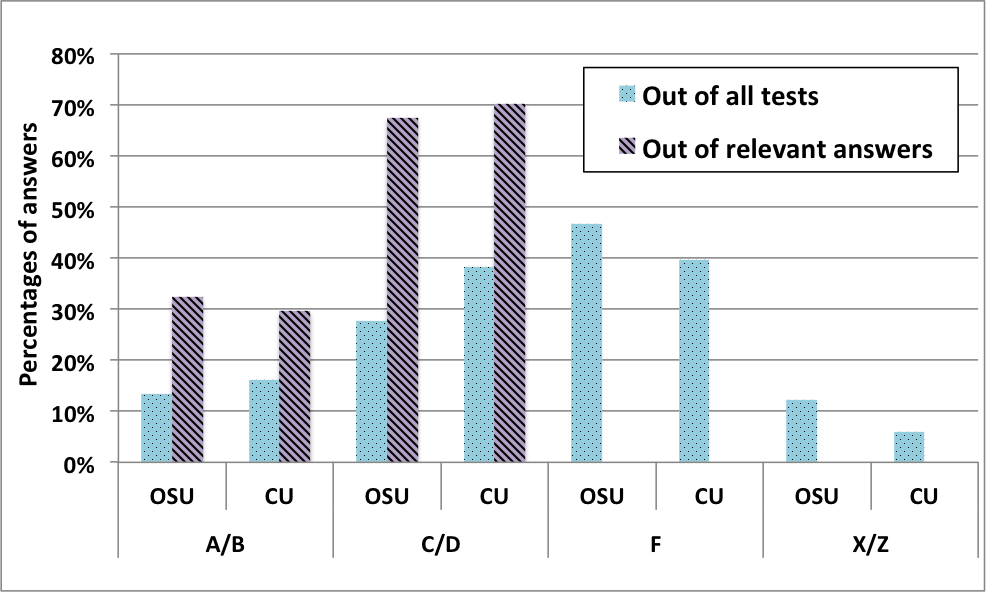}
 \caption{Frequency of correct (A/B), incorrect (C/D), irrelevant (F) and lack of answer (X/Z) at OSU and CU.}
\label{fig:q5data-all}
\end{figure}
\section{Results from a new analysis}
With our new categorization (Table~\ref{tab:newq2/q5}), we compared responses on the superposition question (Fig.~\ref{fig:q5}) for $N^{total}_{OSU}=90$ tests from OSU students and $N^{total}_{CU}=64$ tests provided to us by CU. 

Initially, we only considered answers which were relevant to the problem, \textit{i.e.,} we eliminated F, X, and Z responses. This left $N_{OSU}=37$ and $N_{CU}=37$ students who tried to add/superimpose something (either correctly or incorrectly). Figure~\ref{fig:q5data-rel} shows the distribution of correct answers, between the electric field approach (A) and the potential  approach (B), and incorrect answers, between clearly talking about adding charges (C) and being ambiguous about what should be superposed (D). 

\begin{table}
\begin{tabular}{p{0.25cm}p{0.5cm}p{5.5cm}}
\hline
A&\multicolumn{2}{l}{Clearly talks about adding electric fields}\\
&A1&uses the word ``superposition''\\
&A2&does not use the word ``superposition''\\
\hline
B&\multicolumn{2}{l}{Clearly talks about adding potentials}\\
&B1&uses the word ``superposition''\\
&B2&does not use the word ``superposition''\\
\hline
C&\multicolumn{2}{l}{Seems to be adding charges}\\
&C1&uses the word ``superposition''\\
&C2&does not use the word ``superposition''\\
\hline
D&\multicolumn{2}{l}{Ambiguous about what is being added/superposed}\\
&D1&uses the word ``superposition''\\
&D2&does not use the word ``superposition''\\
\hline
F&\multicolumn{2}{l}{Irrelevant answer}\\
\hline
X&\multicolumn{2}{l}{Did not answer}\\
\hline
Z&\multicolumn{2}{l}{Answered ``I don't know''}\\
\hline
\end{tabular}
\caption{Categories of responses for our analysis.}
\label{tab:newq2/q5}
\end{table}

The first thing to note is that of the relevant responses, only 3\%  of CU students used a potential approach (9\% of correct answers), in contrast to 11\% of OSU students (33\% of correct answers). Even more striking is the difference in the explicit use of the word ``superposition.''  Of all relevant answers (combining A, B, C, D),  $81\%$ of CU students explicitly used the term ``superposition,'' compared to the same percentage of OSU students who did {\bf not} use this term. This pattern is also evident in the correct responses (A and B only). Of all correct answers (combining A and B), $77\%$ of OSU students  {\bf did not} explicitly use the term ``superposition,'' compared to 82\% of CU students who {\bf did} use this term.

In order to look more closely at the issue of {\em what} is being superposed, we also did a comparison without considering the use of the word ``superposition'' or distinguishing between electric field and potential approaches. These results are presented in Fig.~\ref{fig:q5data-all}, which groups all correct categories (A and B) and all incorrect or ambiguous categories (C and D).
The overall results are comparable for both universities. It was surprising to us that in both schools only $\sim15\%$ of all students took a clearly correct (electric or potential field) approach to this problem ($\sim 30\%$ of relevant responses). If we look only at relevant answers, in almost  $70\%$ of cases students were either unclear about what they wanted to add/superimpose or were clearly talking about adding charges.


\section{Discussion}

In most E\&M I courses, students begin with the electric field approach, and only after they master this concept do they move on to the potential approach. As a consequence, the electric field approach is an intuitive and natural way for those students to tackle problems. This might explain why so few CU students tried to solve this problem using potentials, compared to OSU students, who are introduced to potential before electric field. This is a question we are currently exploring.\cite{GoldtrapThesis}

Regarding the difference in explicit use of the word ``superposition'', the CU course materials,\cite{CUEMI} which include lecture notes, clicker questions, tutorials, \emph{etc.}, seem to strongly emphasize the term ``superposition.'' For example,
in both the instructor and student manuals for the first tutorial, 
 one can find the following note:
\begin{quote} 
A common HW and exam problem asks for the E-field caused by a charged disc. Hint: A disc is the sum of many rings. Did someone say ``superposition''?
\end{quote}
This emphasis is not similarly apparent in the Paradigms materials.\cite{wiki}

Additionally, as one might expect at the institution that developed the CUE, one can see the impact of this test in the reformed course materials, such as clicker questions that are similar to questions on the CUE, in whole or in part. This interaction between the development of the course materials and the development of the assessment is not unexpected, but it is important to consider when extending the assessment beyond the original institution.

\section{Conclusions}

Two of the stated goals of the CUE are to ``serve as a comparative instrument to assess upper-division E\&M courses'' and to help ``education researchers to learn more about specific areas where upper-level students are struggling.''\cite{Chasteen09-CUE2} Using the question on the superposition principle as an example, we identify two ways in which the CUE is not fulfilling these goals as well as it could using the current rubric. 

The emphasis in the rubric on the terminology of ``superposition'' and lack of inclusion of a potential approach represent a potential bias toward the traditional subject order and emphasis that obscures interesting differences between approaches.

Even more striking is the lack of information the current rubric provides about an important area where students seem to be struggling, independent of approach. Both OSU and CU have made major reforms designed to help students understand superposition, though in different ways. Yet, despite these reforms, more than 40\% at each school do not recognize the need for superposition on this problem and of those that do, almost 70\% do not clearly identify what they are superposing.

The categorization we propose here allows one to identify how a teaching scheme may impact responses and even more importantly, illuminates an important issue about student understanding of superposition that is obscured in the original rubric. We suggest that a similar reexamination of the rest of the CUE might provide additional examples where insight on student understanding  would broaden the impact of the CUE as a tool for assessing upper-division E\&M courses. 

The University of Colorado is currently working on a multiple-choice version of the CUE. This provides an excellent opportunity for just such a re-analysis. A recent interview with two OSU undergraduates suggests that the issues raised in this paper are still of concern with the multiple-choice version of the CUE.


\begin{theacknowledgments}
Supported in part by NSF DUE 1023120.  We would like to thank Steve Pollock, Rachel Pepper and Bethany Wilcox for conversations about the design and grading of the CUE and for sharing some of the CU test data. 
\end{theacknowledgments}


\end{document}